\documentclass[10pt]{article}
\usepackage{opex2}
\usepackage{epsfig}
\usepackage{amsmath}
\usepackage{amssymb}

\begin{document}

\title{Effective area of photonic crystal fibers}

\author{Niels Asger Mortensen}
\address{Crystal Fibre A/S, Blokken 84, DK-3460 Birker\o d, Denmark}
\email{nam@crystal-fibre.com\\http://www.crystal-fibre.com}

\begin{abstract}
We consider the effective area $A_{\rm eff}$ of photonic crystal fibers (PCFs) with a triangular air-hole lattice in the cladding. It is first of all an important quantity in the context of non-linearities, but it also has connections to leakage loss, macro-bending loss, and numerical aperture. Single-mode versus multi-mode operation in PCFs can also be studied by comparing effective areas of the different modes. We report extensive numerical studies of PCFs with varying air hole size. Our results can be scaled to a given pitch and thus provide a general map of the effective area. We also use the concept of effective area to calculate the ``phase'' boundary between the regimes with single-mode and multi-mode operation.
\end{abstract}

\section{Introduction}

Photonic crystal fibers (PCF) constitute a new class of optical fibers which has revealed many surprising phenomena and also has a big potential from an application point of view (for recent special issues on photonic crystal fibers, see {\it e.g.} Refs. \cite{recent1,recent2}). Here we consider the type of PCF first studied in Ref.~\cite{knight1996,knight1997errata} which consists of pure silica with a cladding with air-holes of diameter $d$ arranged in a triangular lattice with pitch $\Lambda$, see Fig.~1. For a review of the basic operation we refer to Ref.~\cite{broeng1999}.

The effective area is a quantity of great importance. It was originally introduced as a measure of non-linearities; a low effective area gives a high density of power needed for non-linear effects to be significant \cite{agrawal}. However, the effective area can of course also be related to the spot-size $w$ through $A_{\rm eff}=\pi w^2$, and thus it is also important in the context of confinement loss \cite{white2001}, micro-bending loss \cite{petermann1977}, macro-bending loss \cite{sorensen2002}, splicing loss \cite{marcuse1977}, and numerical aperture \cite{mortensen2002}. 

Strictly endlessly single-mode operation of PCFs \cite{birks1997} is possible for $d<d^* \sim 0.45 \Lambda$ \cite{broeng1999}, but even for larger air holes single-mode operation is possible for wavelengths above a certain cut-off $\lambda^*$. We demonstrate that the effective area of the second-order mode is a useful concept in determining this cut-off for a given hole size $d$. For $d<d^*$ we recover the endlessly single-mode regime with $\lambda^* \rightarrow 0$.

\begin{figure}[h!]
\begin{center}
\epsfig{file=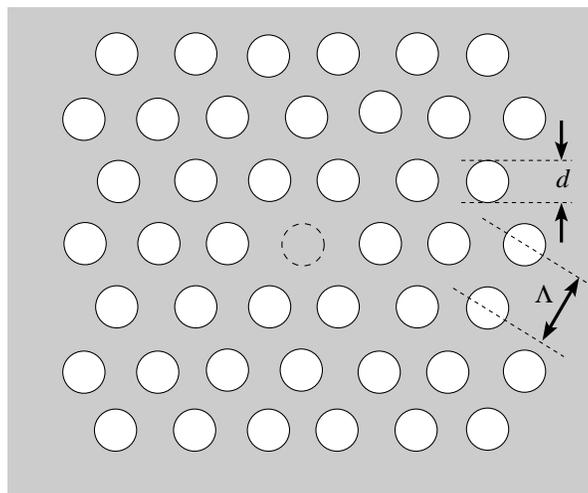, width=0.59\textwidth,clip}
\end{center}
\caption{Schematic of end-facet of an all-silica photonic crystal fiber. The microstructured cladding consists of air holes of diameter $d$ arranged in a triangular lattice with pitch $\Lambda$. The silica core guiding the light is formed by the ``missing'' air hole (indicated by the dashed circle).  }
\label{fig1}
\end{figure}

The paper is organized as follows: In Sec. II we introduce the concept of the effective area and In Sec. III we report numerical results for PCFs of the type shown in Fig.~\ref{fig1}. Finally, discussion and conclusions are given.

\section{Eigen-modes and effective areas}
We consider PCFs realized in dielectric materials with no free charges or currents. From Maxwell's equations it then follows that the ${\boldsymbol H}$-field is governed by the general wave equation (see {\it e.g.} Ref.~\cite{joannopoulos})

\begin{equation}\label{maxwell}
{\boldsymbol \nabla}\times \Big[\varepsilon({\boldsymbol r}){\boldsymbol \nabla}\times {\boldsymbol H}_\omega({\boldsymbol r})\Big]= (\omega/c)^2{\boldsymbol H}_\omega({\boldsymbol r})
\end{equation}
where $\varepsilon$ is the dielectric function and $\omega$ is the frequency of the harmonic mode, $ {\boldsymbol H}_\omega({\boldsymbol r},t)= {\boldsymbol H}_\omega({\boldsymbol r})e^{\pm i\omega t}$.

The PCF has translation symmetry along the fiber axis ($z$-axis) and for $\varepsilon({\boldsymbol r})=\varepsilon({\boldsymbol r}_\perp)$ the solution is of the form 
\begin{equation}
{\boldsymbol H}_\omega({\boldsymbol r})= \sum_n \alpha_n {\boldsymbol h}_n({\boldsymbol r}_\perp)e^{\pm i \beta_n(\omega) z}
\end{equation}
where ${\boldsymbol h}_n$ is the transverse part of the $n$th eigen-mode and $\beta_n$ is the corresponding propagation constant at frequency $\omega$. Often one will specify the free-space wavelength $\lambda$ rather than the frequency $\omega=c (2\pi/\lambda)$.

\begin{figure}[h!]
\begin{center}
\epsfig{file=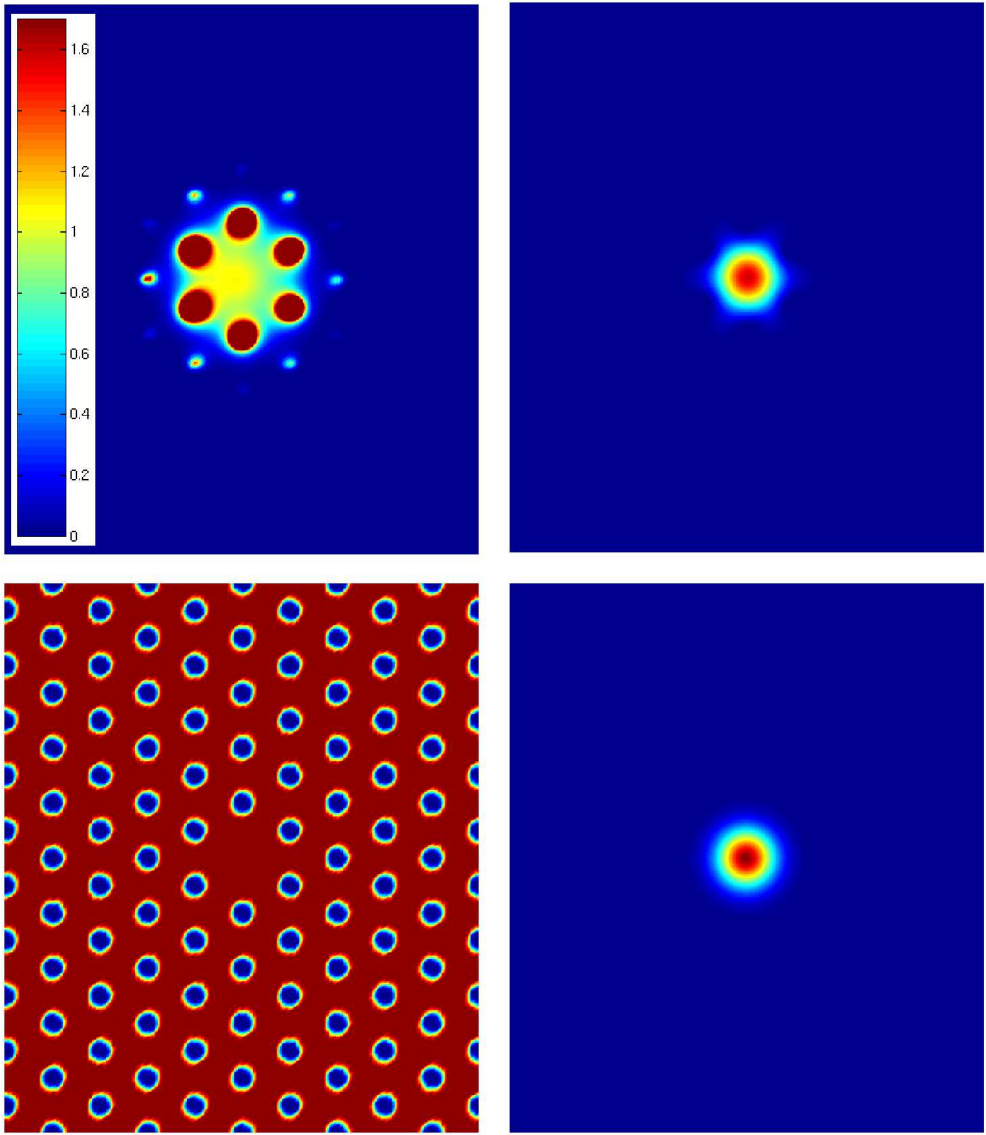, width=0.99\textwidth,clip}
\end{center}
\caption{Comparison of the real intensity $I({\boldsymbol r}_\perp)$ to the Gaussian intensity $I_{G}({\boldsymbol r}_\perp)$ with a width obtained from $A_{\rm eff}$. The upper left panel shows $I_{G}/I$, the upper right panel shows $I$, the lower right panel shows $I_{G}$, and the lower left panel shows the dielectric function $\varepsilon$ with $d/\Lambda=0.3$. The fields are calculated at $\lambda/\Lambda = 0.48$. }
\label{fig2}
\end{figure}

The effective area associated with the $n$th eigen-mode is given by~\cite{agrawal} 

\begin{equation}\label{Aeff}
A_{\rm eff,n}(\lambda)= \Big[\int d{\boldsymbol r}_\perp I_{n}({\boldsymbol r}_\perp)\Big]^2\Big/\int d{\boldsymbol r}_\perp I_{n}^2({\boldsymbol r}_\perp),
\end{equation}
where $I_n({\boldsymbol r}_\perp)=|{\boldsymbol h}_n({\boldsymbol r}_\perp)|^2$ is the intensity distribution.
It is easy to show that for a Gaussian mode ${\boldsymbol h}({\boldsymbol r}_\perp) \propto e^{-(r_\perp/w)^2}$ of width $w$ the effective area is $A_{\rm eff}=\pi w^2$.
Applying Eq.~(\ref{Aeff}) to close-to-Gaussian modes in some way corresponds to a Gaussian fit averaged over all angular directions. In Fig.~\ref{fig2} we illustrate this by an example. For $d/\Lambda=0.3$ and $\lambda/\Lambda = 0.48$ we compare the intensity $I$ of the fundamental mode ($n=1$ or $2$) to the intensity $I_G$ of the corresponding Gaussian with the width calculated from $A_{\rm eff}$. As seen the over-all intensity distribution is reasonably described by the Gaussian with a width calculated from the effective area. 

The low-intensity deviations from a Gaussian intensity distribution are coursed by the six-fold symmetry of the air hole lattice; the field extends slightly into the six silica bridges formed by the six holes nearest to the core. Similar modal properties have been found experimentally, see {\it e.g.} Refs.~\cite{knight1996,eggleton1999,eggleton2000}.

Below we mention a few examples of some of the different phenomena and quantities which can be at least partly quantified from knowledge of the effective area; the non-linearity coefficient, the numerical aperture, the macro-bending loss coefficient, and splicing loss. In Sec. III we also discuss how a second-order mode cut-off can be identified using the concept of the effective area.

\subsection{Non-linearity coefficient}
The non-linearity coefficient $\gamma$ is given by \cite{agrawal}

\begin{equation}
\gamma=\frac{n_2\omega}{c A_{\rm eff}}=\frac{n_2 2\pi}{\lambda A_{\rm eff}}
\end{equation}
where $n_2$ is the nonlinear-index coefficient in the nonlinear part of the refractive index, $\delta n = n_2 | {\boldsymbol E}|^2$. Knowledge of $A_{\rm eff}$ is thus an important starting point in the understanding of non-linear phenomena in PCFs. Due to the high index contrast between silica and air the PCF technology offers the possibility of a much tighter mode confinement (over a wide wavelength range) and thereby a lower effective area compared to standard-fiber technology. Furthermore, the microstructured cladding of the PCFs also allows for zero-dispersion engineering. For a recent demonstration of a highly nonlinear PCF with a zero-dispersion at $\lambda = 1.55\,{\rm \mu m}$ see Ref.~\cite{OFC}.

\subsection{Numerical aperture}

Also the numerical aperture (NA) relates to the effective area. For a Gaussian field of width $w$ one has the standard approximate expression $\tan\theta\simeq \lambda/\pi w$ for the half-divergence angle $\theta$ of the light radiated from the end-facet of the fiber \cite{ghatak1998}. The corresponding numerical aperture can then be expressed as

\begin{equation}\label{ghatak}
{\rm NA}=\sin\theta \simeq  \big(1+ \pi A_{\rm eff}/\lambda^2\big)^{-1/2}.
\end{equation}
In Ref.~\cite{mortensen2002} this was used in a study of the numerical aperture of PCFs.

\subsection{Macro-bending loss}
For estimation of the macro-bending loss coefficient $\alpha$ the Sakai--Kimura formula \cite{sakai1978} can be applied to PCFs. This involves an evaluation of the ratio $A_e^2/P$, where  $A_e$ is the amplitude coefficient of the field in the cladding and $P$ the power carried by the fundamental mode. The Gaussian approximation gives $A_e^2/P=1/A_{\rm eff}$ \cite{sakai1979} and in Ref.~\cite{sorensen2002} this was used to calculate the macro-bending loss in PCFs based on fully-vectorial eigenmodes of Maxwell's equations.

\subsection{Splicing loss}

Also splicing loss can be quantified in terms of the concepts of effective areas. The splicing of two aligned fibers with effective areas $A_{\rm eff,1}$ and $A_{\rm eff,2}$ will have a power transmission coefficient $T<1$ given approximately by

\begin{equation}
T\approx \frac{4A_{\rm eff,1}A_{\rm eff,2} }{(A_{\rm eff,1}+A_{\rm eff,2})^2}
\end{equation}
due to the mismatch of effective areas.

\begin{figure}[h!]
\begin{center}
\epsfig{file=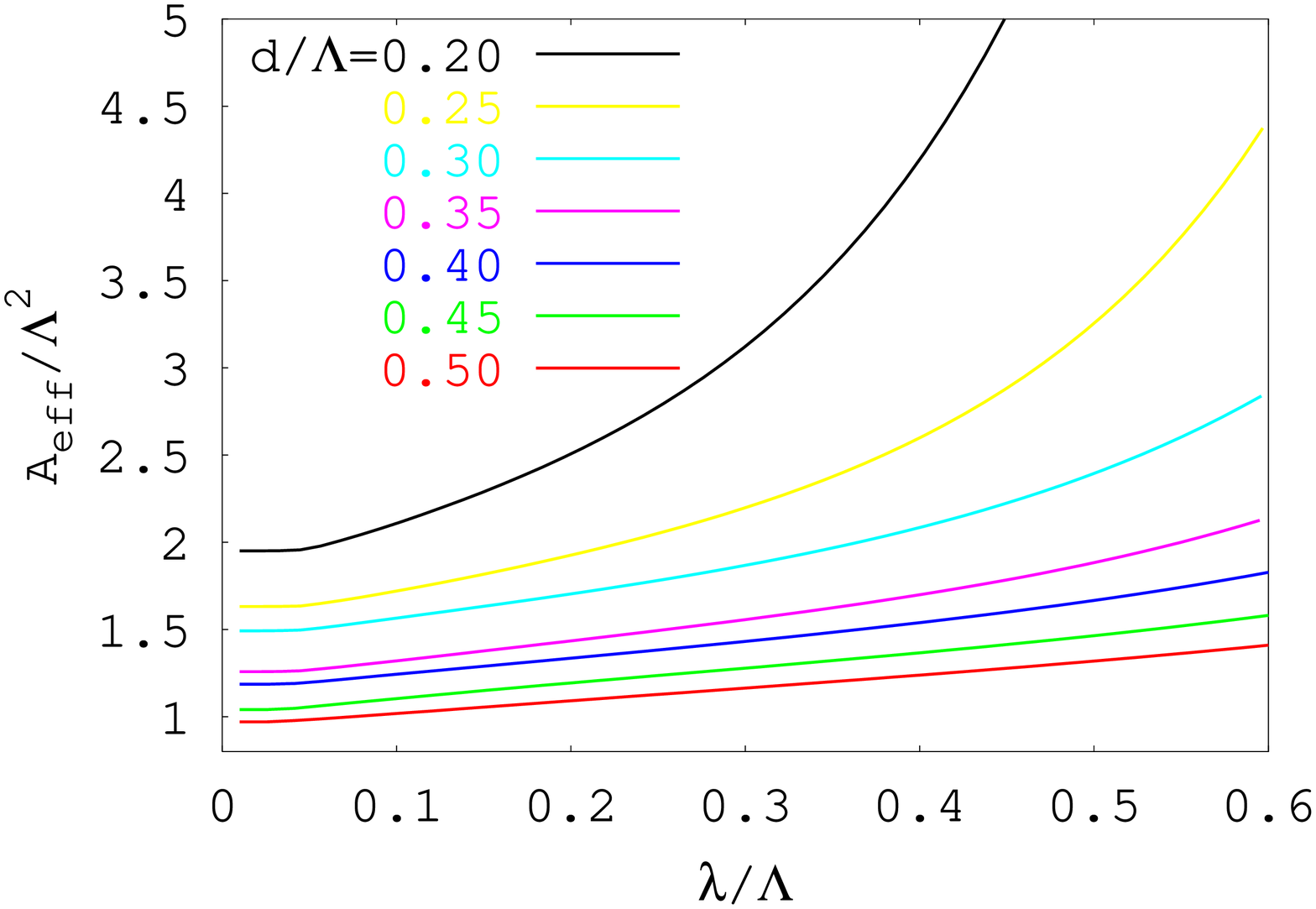, width=0.79\textwidth,clip}
\epsfig{file=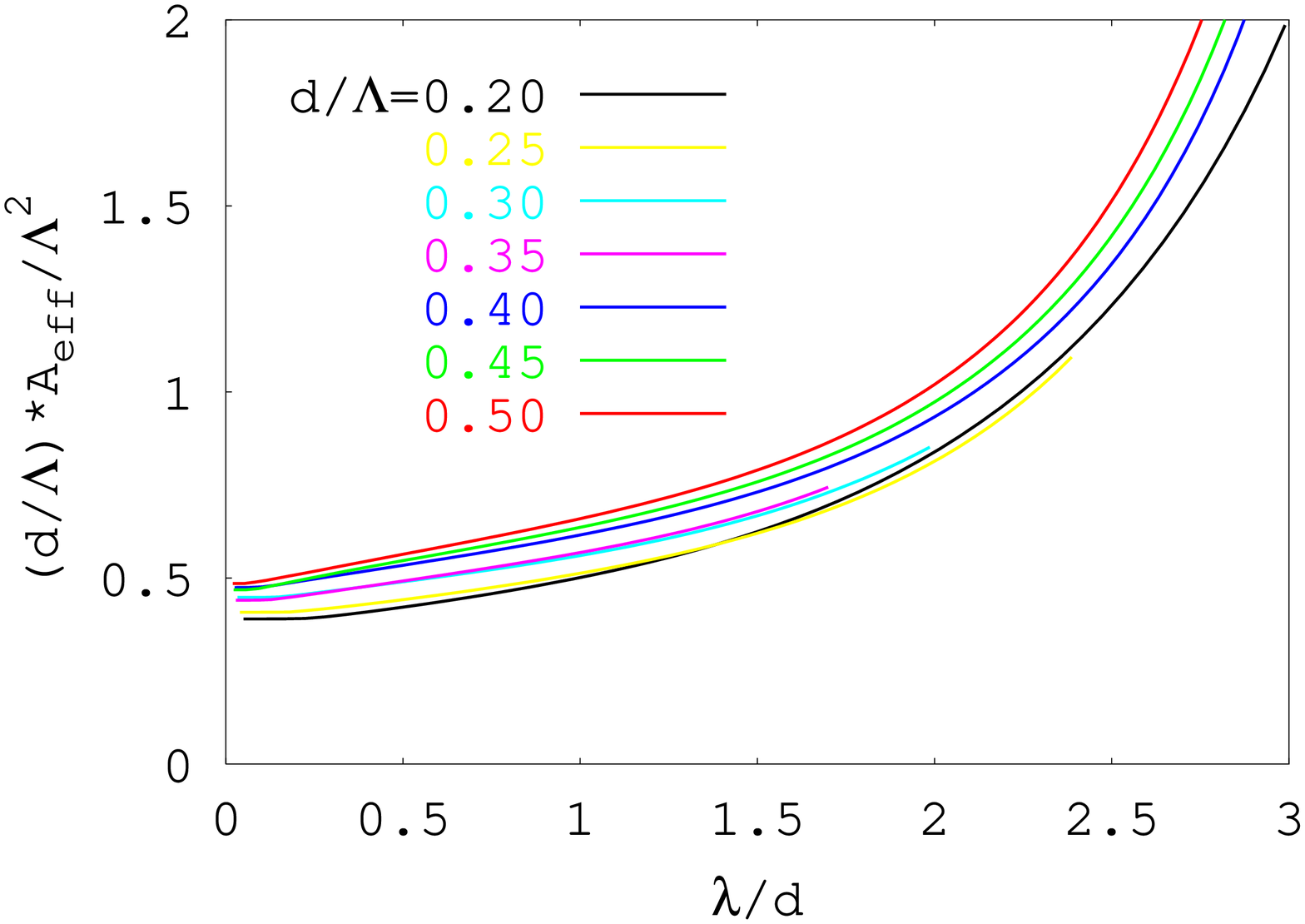, width=0.79\textwidth,clip}
\end{center}
\caption{The upper panel shows the effective area as a function of wavelength for different hole sizes. The lower panel shows the same data as in the upper panel, but with the vertical axis scaled by the factor $d/\Lambda$ and the horizontal axis scaled by the factor $(d/\Lambda)^{-1}$.}
\label{fig3}
\end{figure}

\section{Numerical results}

We solve Eq.~(\ref{maxwell}) numerically and from the eigenmodes we calculate the corresponding effective area from Eq.~(\ref{Aeff}). The fully-vectorial eigenmodes of Maxwell's equations are computed with periodic boundary conditions in a planewave basis \cite{johnson2000}. This approach provides the eigenmodes on a discrete lattice and the integrals in Eq.~(\ref{Aeff}) are then obtained by sums over lattice sites inside the super-cell. For the dielectric function we have used $\varepsilon=1$ for the air holes and $\varepsilon=(1.444)^2 = 2.085$ for the silica. Ignoring the frequency dependence of the latter the wave equation, Eq.~(\ref{maxwell}) becomes scale-invariant \cite{joannopoulos} and all the results to be presented can thus be scaled to the desired value of $\Lambda$.

\subsection{Fundamental mode}

 In Fig.~\ref{fig3} we show the effective area of the fundamental mode as a function of wavelength for different hole sizes. As expected the mode becomes more confined -- lower effective area -- for increasing air hole size. In general we find that the effective area of the fundamental mode is of the order of $\Lambda^2$ with a prefactor which depends slightly on the air hole size. In fact, scaling the results by a factor $d/\Lambda$ we find numerical evidence that

\begin{equation}
A_{\rm eff}\propto (\Lambda/d) \times  \,\Lambda^2 +{\cal O}(\lambda/d), 
 \end{equation} 
and as demonstrated in the lower panel of Fig.~\ref{fig3} the prefactor is of order $0.5$. 

Fig.~\ref{fig3} is a general ``map'' of the effective area which can be used in calculating the different quantities discussed in Sec.~II as well as for designing single-mode PCFs with specified modal properties. If for example the value of $d/\Lambda$ is given (which is often the case during stack-and-pull production of PCFs) a given effective area can be obtained by scaling the fiber structure, {\it i.e.} the pitch $\Lambda$. The Corning SMF28 standard fiber has an effective area $A_{\rm eff}\sim 86\,{\rm \mu m}^2$ at $\lambda = 1550\,{\rm nm}$ and Fig.~\ref{fig3} suggests that a comparable effective area can be realized by the PCF technology using {\it e.g.} $d/\Lambda \sim 0.25$ and $\Lambda \sim 6.5\, {\rm \mu }m$.

\begin{figure}[b!]
\begin{center}
\epsfig{file=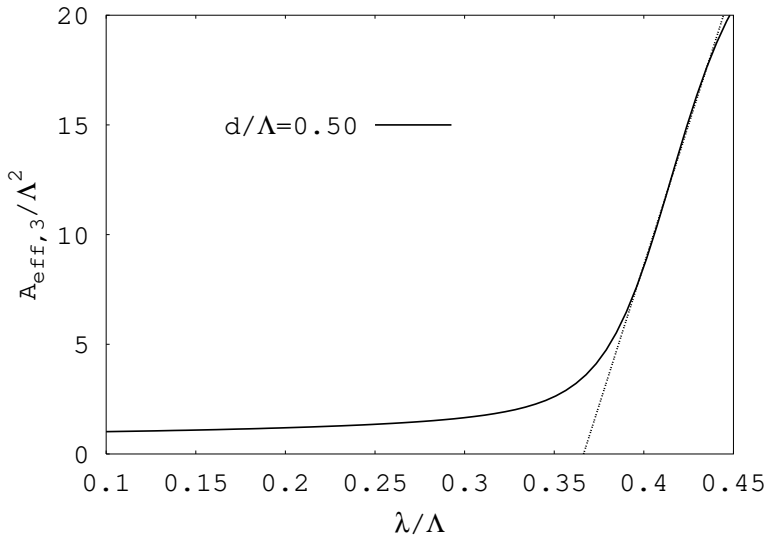, width=0.79\textwidth,clip}
\end{center}
\caption{The effective area of the second-order mode for $d/\Lambda=0.5$. For short wavelengths $A_{\rm eff, 3}\sim \Lambda^2$ and the mode is guided in the core. For high wavelengths the mode becomes a cladding mode with effective area approaching the area of the super-cell used in the calculation. The crossing of the dashed line with the horizontal axis indicates the cut-off for the second order mode.}
\label{fig4}
\end{figure}

\subsection{Second-order mode}

In Fig.~\ref{fig4} we show a calculation of the effective area of the second-order mode ($n=3$) for a PCF with $d/\Lambda=0.5$ air holes. For short wavelengths the effective area is of the order $\Lambda^2$ with the mode confined to the core of the PCF. At long wavelengths the effective area diverges corresponding to a delocalized cladding mode. Even though the transition to a cladding mode is not abrupt the high slope of the effective area allows for a rather accurate introduction of a cut-off wavelength $\lambda^*$. This cut-off wavelength is indicated by the crossing of the dashed line with the horizontal axis. 

\subsection{Single-mode versus multi-mode operation}

Conditions for single-mode operation of PCFs turn out to be very different from standard fiber technology and for not too big air holes, $d < d^*$, they can even be endlessly single-mode \cite{knight1996}. The reason is that the effective cladding index has a wavelength dependence which prevents the core from guidance of higher-order modes while still guiding a fundamental mode. In Ref.~\cite{broeng1999} $d^*\sim 0.45 \Lambda$ was suggested from standard-fiber considerations.

For $d > d^*$ a second-order mode is guided for wavelengths shorter than a certain cut-off and above this cut-off the PCF is single-mode. This cut-off of the second-order mode can be found from studies of the corresponding effective area; below the cut-off the effective area $A_{\rm eff,3}$ is finite and comparable to $\Lambda^2$ and above the cut-off the effective area diverges (in a super-cell calculation it approaches the area of the super-cell).

\begin{figure}[h!]
\begin{center}
\epsfig{file=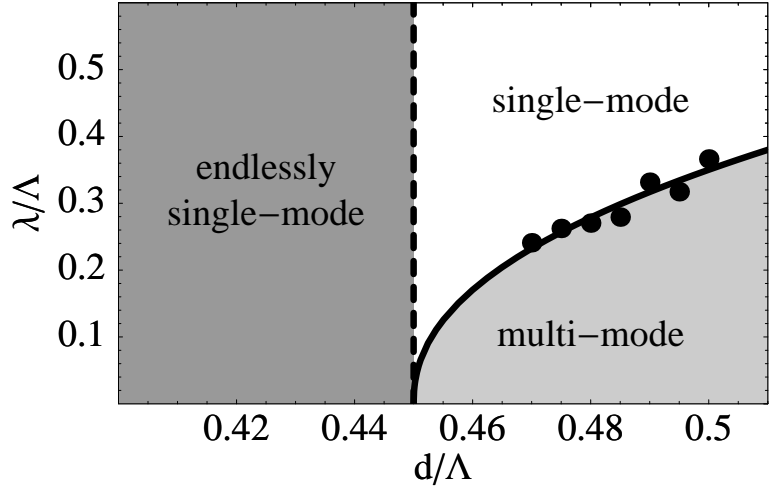, width=0.79\textwidth,clip}
\end{center}
\caption{Diagram illustrating ``phases'' with single-mode and multi-mode operation. The data points are cut-off values obtained from analysis of the form indicated in Fig.~\ref{fig4}.}
\label{fig5}
\end{figure}

Comparing Fig.~\ref{fig4} to Fig.~\ref{fig3} for $d/\Lambda=0.5$ it is seen how we for wavelengths shorter than $\lambda^* \sim 0.35 \Lambda$ have a multi-mode PCF whereas the PCF is in the single-mode regime for $\lambda > \lambda^*$. Carrying out the same analysis for different hole-sizes allows for constructing a ``phase'' diagram, see Fig. \ref{fig5}. The data points indicate calculated cut-off wavelengths and the solid line is a fit to the function

\begin{equation}
f(x)=\alpha(x-x_0)^\beta,\; x_0=0.45,
\end{equation}
with the fitting coefficients $\alpha\simeq 1.34$ and $\beta\simeq 0.45$.

The solid line defines a ``phase'' boundary; above the line the PCF is single-mode and below it is multi-mode. Furthermore there is an endlessly single-mode regime for $d < d^* \sim 0.45\Lambda$ \cite{broeng1999}. In principle the effective area approach allows for an independent determination of $d^*$  but it should be emphasized that the numerical efforts needed to resolve the cut-off increase dramatically when $d$ approaches $d^*$. 

\section{Conclusion}

We have considered the effective area $A_{\rm eff}$ of photonic crystal fibers (PCFs) with a triangular air-hole lattice in the cladding. Based on extensive numerical studies of PCFs with varying air hole size we have constructed a map of the effective area which can be scaled to a desired value of the pitch.
We have also utilized the concept of effective area to calculate the ``phase'' boundary between the regimes with single-mode and multi-mode operation. In this work we have studied PCFs with a triangular air-hole lattice cladding, but we emphasize that the approach applies to microstructured fibers in general.

\section*{Acknowledgement}
The author has benefitted from useful discussions with J.~R. Jensen, P.~M.~W. Skovgaard, M.~D. Nielsen, J. Broeng, and K.~P. Hansen.


\begin{thebibliography}{10}

\bibitem{recent1}
Opt. Express {\bf 9}, 674--779 (2001), http://www.opticsexpress.org/issue.cfm?issue id=124

\bibitem{recent2}
J. Opt. A: Pure Appl. Opt. {\bf 3}, S103--S207 (2001).

\bibitem{knight1996}
J.~C. Knight, T.~A. Birks, P.~S.~J. Russell, and D.~M. Atkin, ``All-silica
  single-mode optical fiber with photonic crystal cladding,'' Opt. Lett. {\bf 21}, 1547--1549 (1996).

\bibitem{knight1997errata}
J.~C. Knight, T.~A. Birks, P.~S.~J. Russell, and D.~M. Atkin, ``All-silica
  single-mode optical fiber with photonic crystal cladding: errata,'' Opt.
  Lett. {\bf 22}, 484--485 (1997).

\bibitem{broeng1999}
J.~Broeng, D.~Mogilevstev, S.~E. Barkou, and A.~Bjarklev, ``Photonic crystal
  fibers: A new class of optical waveguides,'' Opt. Fiber Technol. {\bf 5}, 305--330 (1999).

\bibitem{agrawal}
G.~P. Agrawal, {\em Nonlinear Fiber Optics} (Academic Press, San Diego, 2001).

\bibitem{eggleton1999}
B.~J. Eggleton, P.~S. Westbrook, R.~S. Windeler, S. Sp\"{a}lter, and T.~A. Strasser, ``Grating resonances in air-silica microstructured optical fibers,'' Opt. Lett. {\bf 24}, 1460--1462 (1999).

\bibitem{eggleton2000}
B.~J. Eggleton, P.~S. Westbrook, C.~A. White, C. Kerbage, R.~S. Windeler, G.~L. Burdge, ``Cladding-Mode-Resonances in Air-Silica Microcstructed Optical Fibers,'' J. Lightwave Technol. {\bf 18}, 1084--1100 (2000).

\bibitem{OFC}
K.~P. Hansen, J.~R. Jensen, C. Jacobsen, H.~R. Simonsen, J. Broeng, P.~M.~W. Skovgaard, A. Petersson, and A. Bjarklev, ``Highly Nonlinear Photonic Crystal Fiber with Zero-Dispersion at 1.55 $\mu$m'', OFC 2002 Postdeadline Paper, FA9-1.

\bibitem{white2001}
T.~P. White, R.~C. McPhedran, C.~M. {de Sterke}, L.~C. Botton, and M.~J. Steel,
  ``Confinement losses in microstructured optical fibers,'' Opt. Lett. {\bf 26}, 1660--1662 (2001).

\bibitem{petermann1977}
K.~Petermann, ``Fundamental mode microbending loss in graded index and w
  fibers,'' Opt. Quantum Electron. {\bf 9}, 167--175 (1977).

\bibitem{sorensen2002}
T.~S{\o}rensen, N.~A. Mortensen, J.~Broeng, A.~Bjarklev, T.~P. Hansen,
  E.~Knudsen, S.~E.~B. Libori, H.~R. Simonsen, and J.~R. Jensen, ``Spectral
  macro-bending loss considerations on photonic crystal fibres,'' IEE
  Proc.-Optoelectron., submitted.

\bibitem{marcuse1977}
D.~Marcuse, ``Loss analysis of sigle-mode fiber splices,'' Bell Syst.
  Tech. J. {\bf 56}, 703 (1977).

\bibitem{mortensen2002}
N.~A. Mortensen, J.~R. Jensen, P.~M.~W. Skovgaard, and J.~Broeng, ``Numerical
  aperture of single-mode photonic crystal fibers,'' preprint, http://arxiv.org/abs/physics/0202073

\bibitem{birks1997}
T.~A. Birks, J.~C. Knight, and P.~S.~J. Russell, ``Endlessly single mode
  photonic crystal fibre,'' Opt. Lett. {\bf 22}, 961--963 (1997).

\bibitem{joannopoulos}
J.~D. Joannopoulos, R.~D. Meade, and J.~N. Winn, {\em Photonic crystals:
  molding the flow of light} (Princeton University Press, Princeton, 1995).

\bibitem{ghatak1998}
A.~K. Ghatak and K.~Thyagarajan, {\em Introduction to Fiber Optics} (Cambridge University Press, Cambridge, 1998).

\bibitem{sakai1978}
J.~Sakai and T.~Kimura, ``Bending loss of propagation modes in arbitrary-index
  profile optical fibers,'' Appl. Optics {\bf 17}, 1499--1506 (1978).

\bibitem{sakai1979}
J.~Sakai, ``Simplified bending loss formula for single-mode optical fibers,'' Appl. Optics {\bf 18}, 951--952 (1979).

\bibitem{johnson2000}
S.~G. Johnson and J.~D. Joannopoulos, ``Block-iterative frequency-domain
  methods for \uppercase{M}axwell's equations in a planewave basis,'' Opt.
  Express {\bf 8}, 173--190 (2000), http://www.opticsexpress.org/abstract.cfm?URI=OPEX-8-3-173


\end{thebibliography}
\end{document}